\documentclass{article}

\usepackage{graphicx}

\newcommand{\beq}{\begin{equation}}
\newcommand{\eeq}{\end{equation}}
\newcommand{\beqa}{\begin{eqnarray}}
\newcommand{\eeqa}{\end{eqnarray}}
\newcommand{\sr}{\sqrt}
\newcommand{\fr}{\frac}
\newcommand{\mn}{\mu \nu}

\newcommand{\vp}{\varphi}

\def\abstract#1{\vskip 7mm
        \begin{center}{\large Abstract}\par \smallskip
                \begin{minipage}[c]{12cm}
                        \small #1
                \end{minipage}
        \end{center}
}
\def\title#1{\begin{center}{\Large\bf #1}\end{center}}
\def\author#1{\vskip 5mm \begin{center}{#1}\end{center}}
\def\address#1{\begin{center}{\it #1}\end{center}}
\makeatletter
\@ifundefined{lesssim}{}{}
\@ifundefined{gtrsim}{}{}
\def\vereq#1#2{\lower3pt\vbox{\baselineskip1.5pt \lineskip1.5pt
\ialign{$\m@th#1\hfill##\hfil$\crcr#2\crcr\sim\crcr}}}
\makeatother

\begin{document}

\title{%
Stability of smeared black branes and the Gubser-Mitra conjecture
}
\author{%
  Gungwon Kang\footnote{E-mail:gwkang@kias.re.kr}
  and
  Youngone Lee\footnote{E-mail:youngone@phya.yonsei.ac.kr}
}
\address{%
  $^1$School of Physics, Korea Institute for Advanced Study (KIAS),
  207-43 Cheongnyangni 2-dong, Dongdaemun-gu, Seoul 130-722, Korea\\
  $^2$Department of Physics, Yonsei University, 134 Shinchon
  Seodaemoon-gu, Seoul 120-749, Korea
}

\abstract{
Recently, it was suggested that the system of smeared
black branes might provide a counter example to the Gubser-Mitra
conjecture. Concerning to this issue, we have investigated the
$s$-wave perturbation analysis to see how the stability of such
system behaves. Some partial results are reported in this meeting.
}

\section{Introduction}

It is known that a neutral static black string is unstable under
small perturbations - the so-called Gregory-Laflamme instability
\cite{Gregory:1993vy}. When other physical quantities such as
charge, angular momentum and cosmological constant are added,
however, its stability behavior becomes very diverse, depending on
the values of those physical quantities and gravity theories
\cite{Hirayama:2001bi,Hirayama:2002hn}. For instance, a certain
class of charged black branes become stable as the amount of
charge increases \cite{Hirayama:2002hn}. Concerning such change of
stability behavior, Guber and Mitra conjectured that a noncompact
uniform black brane becomes stable if and only if it becomes
locally thermodynamically stable \cite{Gubser:2000mm}. This
Gubser-Mitra (GM) conjecture has been checked for various black
brane solutions, and seems to hold well so far
\cite{Hirayama:2002hn}.
\\
Recently, however, Bostock and Ross suggested that the GM
conjecture might not hold for a certain black brane background
having smeared electric charges along a circle
\cite{Bostock:2004mg}. In more detail, let us consider a gravity
theory with a dilaton and a $(p+2)$-form field strength whose
action is given by
 \beq
 I = \fr{1}{16\pi G_D} \int d^Dx \sr{-g} \left[ R -\fr{1}{2}
 \partial_M \phi \partial^M \phi -\fr{1}{2(p+2)\!} e^{a\phi}
 F^2_{(p+2)} \right] .
 \eeq
The metric for electrically charged dilatonic black $p$-brane
solutions smeared on transverse circle is\footnote{The case of
$\beta =1$ ({\it i.e.}, $a^2= 4-2(D-d)(d-2)/(D-2)$) corresponds to
the one that Bostock and Ross considered. Here we give solutions
for arbitrary values of $a$ since we will consider the case of
$a=0$ in the perturbation analysis below for simplicity.}
 \beqa
 \mbox{} && ds^2 = H^{-\fr{d-2}{D-2}} \Bigg[ -f dt^2
+H\Big(\fr{dr^2}{f} +r^2 d\Omega^2_{d-2} \Big) +Hdz^2 +d\vec{x}^2
\Bigg]  \nonumber \\
&& e^{2\phi} = \beta^{-2/a} H^a, \quad A_{01\cdots p} = \coth \mu
(1-H^{-1/\beta}), \quad f=1-\Big(\fr{r_0}{r}\Big)^{d-3}, \nonumber
\\
&& H= \Big[ 1+\Big(\fr{r_0}{r}\Big)^{d-3} \sinh^2 \mu
\Big]^{\beta}, \quad \beta=\fr{4}{a^2 +2(D-d)(d-2)/(D-2)} .
 \label{smearedBB}
 \eeqa
The mass and electric charge densities are given by
 \beq
M \sim r_0^{d-3} \left[ (d-2) +(d-3)\sinh^2\mu \right], \qquad Q
\sim r_0^{d-3} \sinh 2\mu ,
 \eeq
respectively. Bostock and Ross' claim is based on several
interesting observations. As usual the stability of the black
brane background above may be determined by checking whether or
not there exists any regular threshold mode in its $s$-wave
perturbation analysis. The threshold mode is a linearized solution
which is static. On the other hand, Harmark and Obers (HO)
introduced an ansatz for any static black brane solution that is
non-uniform in general, but still spherically symmetric in the
direction perpendicular to the world volume \cite{Harmark:2002tr}.
In this ansatz the dilaton and gauge fields are given as in
Eq.(\ref{smearedBB}), and the metric has two unknown functions.
Interestingly, it turns out that the equations of motion for these
functions are independent of the `charge' parameter $\mu$. Thus, a
solution for the uncharged case ({\it i.e.}, $\mu =0$) gives rise
to a solution for any value of non-vanishing $\mu$ as well. Since
the linearized version of these field equations would have the
same property of $\mu$-independence, it presumably implies that
the threshold unstable mode which is known to exist in the
uncharged case must persist to exist for any value of $\mu$ with
given $r_0$, indicating the appearance of dynamical instability in
any charged black brane background in Eq.(\ref{smearedBB}).
However, regarding a segment of this black brane as a thermal
system in the context of black hole thermodynamics, one can easily
find that it becomes thermodynamically stable as the charge
increases ({\it i.e.}, for $\sinh^2\mu > \sinh^2 \mu_{\rm cr}
=1/(d-5)$). Therefore, Bostock and Ross suggested in
Ref.\cite{Bostock:2004mg} that the system of charged black branes
smeared on a transverse circle in Eq.(\ref{smearedBB}) provides a
counter-example to the GM conjecture.
\\
Thus, it is of interest to explicitly perform the linearized
stability analysis for smeared charged black branes in order to
see if such claim is indeed correct. The extremal smeared black
brane is defined as $r_0 \rightarrow 0$ and $\mu \rightarrow
\infty$ with $r_0^{d-3}e^{2\mu}$ fixed, and is believed to be
stable since it corresponds to a BPS state. However, the
independence of the parameter $\mu$ in field equations implies
that the threshold Kaluza-Klein (KK) mass behaves like
$m^*(r_0,\mu) = m^*_0/r_0$ where $m^*_0$ is a constant without
depending on $\mu$. Hence the threshold mass $m^*$ is independent
of the 'charge' parameter. Consequently, the threshold mass does
not seem to decrease down to zero, but seems to diverge in the
extremal limit. Our stability analysis that is not based on the HO
ansatz may also clarify this strange behavior.

\section{$s$-wave perturbation analysis}

\begin{figure}
\setlength{\unitlength}{1cm}
\begin{picture}(15,10)
\put(1,0.1){\Large{0.0}} \put(14.5,0.1){\Large{b}}
\put(0.3,7.7){\Large{$\bar{m}$}}
\includegraphics[width=14.5cm,height=8cm] 
{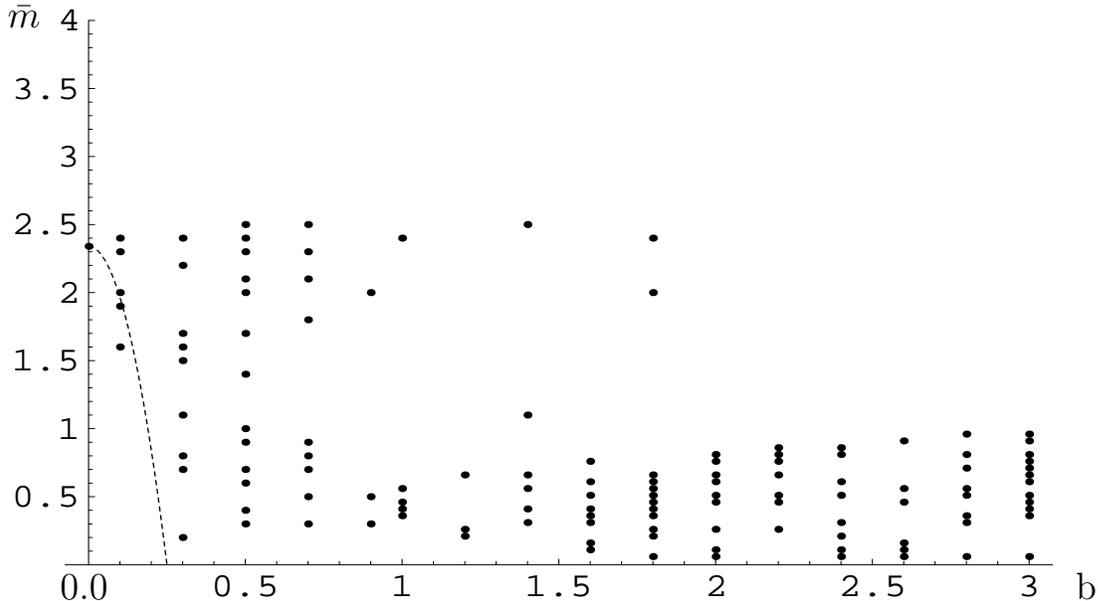}
\end{picture}
\caption{Threshold masses for varying $b=\sinh^2\mu$}
\label{figCGM}
\end{figure}

For simplicity we consider the case of $D=10$ and $a=p=0$. Then
the linearized perturbation equations for $\delta g_{MN}=h_{MN}$,
$\delta F_{MN}$, and $\delta \phi$ become
 \beqa
\mbox{} && \nabla^2 h^M_N +\nabla^M\nabla_N h
-(\nabla_Q\nabla^Mh^Q_N +\nabla_Q\nabla_Nh^{MQ}) +g^{MP}({F_N}^Q
\delta F_{PQ}+{F_P}^Q \delta F_{NQ})  \nonumber
\\
&& \qquad\qquad\qquad\qquad\quad -F^{MP}F_{NQ}h^Q_P -\fr{1}{16}
\left[ F^2 h^M_N +\delta^M_N \left( 2F^{PQ}\delta F_{PQ}
-2h^P_QF_{PR}F^{QR} \right) \right] =0,
\label{leqg} \\
&& \nabla_M\delta F^{MN} -\nabla_MF^{PN} h^M_P -F^{MP}
\nabla_Mh^N_P -\nabla_M(h^M_P -\fr{1}{2} h\delta^M_P ) F^{PN} =0,
\quad \nabla^2 \delta \phi =0. \label{leqF}
 \eeqa
Note that the equation for the dilaton fluctuation is decoupled
and so one can set $\delta \phi =0$. It should be pointed out that
$\delta {\bf F} =\delta \phi =0$ in the analysis based on the HO
ansatz. For threshold modes in $s$-wave fluctuations, one may
assume
 \beqa
\mbox{} && h^M_N= e^{imz} {\rm diag} (\vp(r),\psi(r),\chi(r),
\cdots,\chi(r),0) ,  \nonumber \\
&& \delta {\bf F} = e^{imz} \left( A(r) dr \wedge dt +B(r) dr
\wedge dz +C(r) dt \wedge dz +D(r) d\theta^1 \wedge \cdots \wedge
d\theta^{d-2} \right).
 \eeqa
The Bianchi identity ({\it i.e.}, $d\delta {\bf F}=0$) gives
$D(r)=0$ and $A(r)=i m^{-1}C'(r)$. Both the $h^t_t$-component in
Eq.(\ref{leqg}) and the $r$-component in Eq.(\ref{leqF}) give
$B(r)=0$. Thus, it leads to four second-order and two first-order
coupled ordinary differential equations for four unknown functions
$\vp(r), \psi(r), \chi(r)$, and $C(r)=imP(r)$.
\\
Further reduction finally gives
 \beqa
\mbox{} && \vp'' +\fr{7r^6+2r_0^6}{r(r^6-r_0^6)}\vp'
-\fr{8m^2r^6}{r^6-r_0^6}\vp -\fr{3}{r(r^6-r_0^6)}\psi'
-\fr{7m^2r^6}{r^6-r_0^6}\psi
+\fr{21}{r(r^6-r_0^6)}\chi' -\fr{49m^2r^6}{r^6-r_0^6}\chi =0 ,
\label{leqvp} \\
&& \chi'' +\fr{2(7r^6-4r_0^6)}{r(r^6-r_0^6)}\chi'
+\fr{6(m^2r^6+2r^4)}{r^6-r_0^6}\chi +\fr{1}{r}\vp'
+\fr{m^2r^6}{r^6-r_0^6}\vp -\fr{1}{r}\psi'
+\fr{m^2r^6-12r^4}{r^6-r_0^6}\psi =0 ,
\label{leqchi} \\
&& -\fr{2(49r^6+25r_0^6b)}{7r(r^6+r_0^6b)}\vp'
-\fr{6m^2r^6}{r^6-r_0^6}\vp +\fr{48b}{7r(r^6+r_0^6b)}\psi'
-\fr{4(2m^2r^6-21r^4)}{r^6-r_0^6}\psi
\nonumber \\
&& \qquad\qquad\qquad\qquad\qquad\qquad
-\fr{6[14r^{12}+(22b-7)r_0^6r^6-15r_0^{12}b]}{r(r^6-r_0^6)(r^6+r_0^6b)}\chi'
-\fr{42(m^2r^6+2r^4)}{r^6-r_0^6}\chi =0,
\label{leqpsi}
 \eeqa
where $b=\sinh^2\mu$. Note that, as can be expected from the form
of background fields in Eq.(\ref{smearedBB}), the $r_0$ dependence
in perturbation equations above can completely be eliminated by
rescaling $r$ and $m$ as $\bar{r}=r/r_0$ and $\bar{m}=mr_0$,
respectively.\footnote{In addition, we have $\bar{z}=z/r_0$ and
$\bar{t}=t/r_0$.}
\\
By using Mathematica, for a given value of the `charge' parameter
$b$ we have numerically searched for the value of $\bar{m}$ which
allows a mode solution of these coupled equations that is regular
between the horizon ({\it i.e.}, $\bar{r}\sim 1$) and the spatial
infinity ({\it i.e.}, $\bar{r}\sim \infty$). If there exists no
such non-vanishing value of $\bar{m}$, it indicates at least no
$s$-wave mode instability in the backgrounds corresponding to that
given value of $b=\sinh^2\mu$ and an arbitrary horizon radius of
$r_0 (>0)$. Our numerical results obtained are shown in
Fig.\ref{figCGM}. This numerical search has been performed for
various values of $b$ up to $b=3.0$. Since the extremality
parameter for a given mass density $M$ may be defined as
$q=Q/Q_{\rm max}= (d-3)\sr{b(1+b)}/(d-2+(d-3)b)$, this value of
$b=3.0$ corresponds to $q \simeq 0.83$ which is somewhat close to
the extremal point $q=1$.

\noindent For the uncharged case ({\it i.e.}, $b=0$) three coupled
equations reduce to a single equation, and numerical calculations
show that a regular solution exists when the KK mass parameter is
around $\bar{m} \simeq 2.34$. The heat capacity changes its sign
from $-$ to $+$ at $b=b_{\rm cr}=1/4$. Thus, if the GM conjecture
still holds for this system of smeared black branes, it predicts
that the value of $\bar{m}$ decreases down to zero when $b$
approaches to $b_{\rm cr}$ as indicated in Fig.\ref{figCGM} by a
dotted curve. As can be seen in Fig.\ref{figCGM}, our numerical
results do not seem to follow such behavior. Instead, it seems
that unstable modes still persists to exist even for $b > b_{\rm
cr}$, at least up to $b=3$. However, our results presumably
contain lots of numerical error since there appear multiple points
for a given value of $b$.

\section{Discussion}

To conclude we have considered $s$-wave linearized stability
analysis for electrically charged smeared black brane backgrounds
without using the HO ansatz. It is seemingly that this class of
black branes provides a counter example to the GM conjecture as
suggested by Bostock and Ross in Ref.\cite{Bostock:2004mg}.
However, it is necessary to reduce numerical error in order for
our result to be more conclusive.
\\
It should be pointed out that Eq.(\ref{leqpsi}) depends on the
`charge' parameter $b$ although Eqs.(\ref{leqvp}) and
(\ref{leqchi}) are independent of $b$. Hence threshold masses
$\bar{m}$ allowing regular mode solutions are generically expected
to depend on the value of the `charge' parameter $b$, which is in
disagreement with the expectation in Bostock and Ross' analysis
based on the HO ansatz. For the extremal black brane $r_0
\rightarrow 0$ and $b \rightarrow \infty$ with $b r_0^6$ fixed.
Since $m(r_0,b)=\bar{m}(b)/r_0$, the threshold mass $m$ will
diverge in the extremal limit unless $\bar{m}$ decreases faster
than $b^{-1/6}$. Although it is not definite, $\bar{m}$ in
Fig.\ref{figCGM} does not seem to decrease much down to zero as
$b$ increases, implying that the threshold mass probably diverges
in the extremal limit. This singular behavior may be consistent
with the fact that curvature at the horizon which is given by
 \beq
R_{MNPQ}R^{MNPQ} = \fr{48(2303-5852 b+7937 b^2)}{49(1+b)^{18/7}
r_0^4}
 \eeq
diverges as the smeared black brane in consideration approaches
the extremal point. In Bostock and Ross' analysis the similar
asymptotic behavior of $m$ is expected since $\bar{m}$ is simply a
non-vanishing constant, independent of $b$. Thus, it is very
interesting to understand how this feature is consistent with the
expectation that the extremal smeared black brane is stable under
small perturbations.
\\
Finally, let us speculate why the GM conjecture may not work for
this system of electrically charged smeared black branes in the
context of Reall's proof for the conjecture in
Ref.\cite{Reall:2001ag}. Ignoring the overall conformal factor,
one can express the metric of the smeared black string considered
at the present work as follows;
 \beq
ds^2_{\rm BS} \sim ds^2_{\rm BH} +H(r) dz^2 .
\label{BSReall}
 \eeq
Note that $H(r)=1$ for the class of charged black brane
backgrounds in Reall's partial proof of the conjecture
\cite{Reall:2001ag}. The $s$-wave perturbation analysis for the
metric in the form of Eq.(\ref{BSReall}) gives
 \beq
\Delta_L H_{\mu\nu} = \lambda H_{\mn}
-\fr{1}{4}g^{rr}H'(\partial_r +\cdots )H_{\mn}
-\fr{1}{8}g^{rr}\fr{H'^2}{H}(\delta^r_{\mu}H_{r\nu}+\delta^r_{\nu}H_{r\mu})
\equiv \lambda_{\rm eff} H_{\mn},
 \eeq
where $\Delta_L$ is the Euclidean Lichnerowicz operator for the
black hole part in Eq.(\ref{BSReall}), $H_{\mn}$ the metric
perturbation of that black hole, and $\lambda =-m^2/2$. As was
shown in Reall's proof, the sign of the heat capacity of the black
hole system can be related to the sign of the Euclidean action
integral $\int H^{\mn}\Delta_LH_{\mn} \sim \lambda_{\rm eff}$. The
existence of classical dynamical instability implies that there is
indeed a threshold mode, {\it i.e.}, a non-vanishing $m$ and so
negative $\lambda$. Therefore, it might be possible for $b>b_{\rm
cr}$ that the heat capacity is positive ({\it i.e.},$\lambda _{\rm
eff} >0$) whereas a black brane is still unstable ({\it i.e.},
$\lambda <0$). That is, such violation of the GM conjecture might
occur due to the non-flatness in the smeared direction. Further
investigations for this interesting issue are required in more
detail.

\vspace{1cm} \noindent We find the work \cite{Ross:2005vh} in
which it is argued that the GM conjecture continues to hold for
the cases considered at the present paper, contrary to the claims
of Ref.\cite{Bostock:2004mg} and our present work.

\end{document}